\documentclass[10pt]{article}

\usepackage{times,mathptmx}

\usepackage{geometry}
\usepackage{amsmath,amsthm,amssymb}

\usepackage[font=small]{caption}

\usepackage{graphics}
\usepackage{graphicx}
\usepackage{color}

\usepackage[compact]{titlesec}
\titleformat{\section}{\large\bfseries}{\thesection}{1em}{}
\titleformat{\subsection}{\normalsize\bfseries}{\thesubsection}{1em}{}
\titlespacing*{\section}{0em}{1em}{0.2em}
\titlespacing*{\subsection}{0em}{0.8em}{0.2em}
\titlespacing*{\subsubsection}{0em}{0.8em}{0.2em}

\usepackage{fancyhdr}
\usepackage[square,semicolon]{natbib}
\renewcommand{\cite}[1]{\citep{#1}} 
\usepackage[T1]{fontenc}   
\usepackage[utf8]{inputenc} 

\usepackage{enumitem}
\setlist[enumerate]{itemsep=0mm}

\makeatletter
\renewcommand{\@biblabel}[1]{}
\makeatother

\renewcommand{\title}[1]{
  \begin{center}
    {\LARGE \textbf{#1}}
  \end{center}
}

\makeatletter
\newcommand{\authorinfo}[2]{
  \ifdefined \authorinfos
    \protected@edef \authorinfos{\authorinfos; \textsuperscript{#2}#1}
  \else
    \protected@edef \authorinfos{\textsuperscript{#2}#1}
  \fi
}
\renewcommand{\author}[2]{
  \ifdefined \authors
    \protected@edef \authors{\authors, #1\textsuperscript{#2}}
  \else
    \protected@edef \authors{#1\textsuperscript{#2}}
  \fi
}
\makeatother
\newcommand{\printauthors}{
  \begin{center}
    \noindent {\large \textbf{\authors}} \\
    \textit{\authorinfos}
  \end{center}
}

\newcommand{\correspondence}[2]{
  \begin{center}
    {\small Correspondence: #1. E-mail:~#2}
  \end{center}
}

\renewcommand{\abstract}[2]{
  \hrule
  \vspace{0.6em}
  \noindent \textbf{\textit{Abstract.}} #1 \par
  \vspace{1em}
  \noindent {\small\textit{Keywords:} #2}
  \vspace{0.6em}
  \hrule
  \vspace{2em}
}

\begin{document}

\title{Active Inference for Adaptive BCI: application to the P300 Speller}

\authorinfo{Inria, Bordeaux, France}{1}
\authorinfo{CRNL Inserm, CNRS, Lyon, France}{2}
\authorinfo{Ullo, La Rochelle, France}{3}
\authorinfo{GATE,~CNRS,~Lyon,~France}{4}

\author{Jelena~Mladenović}{1,2}
\author{Jérémy~Frey}{3}
\author{Emmanuel~Maby}{2}
\author{Mateus~Joffily}{4}
\author{Fabien~Lotte}{1}
\author{Jérémie~Mattout}{2}

\printauthors

\correspondence{Jelena Mladenović, Inria, Bordeaux, France}{jelena.mladenovic@inria.fr}

\abstract{

Adaptive Brain-Computer interfaces (BCIs) have shown to improve performance, however a general and flexible framework to implement adaptive features is still lacking. We appeal to a generic Bayesian approach, called Active Inference (AI), to infer user's intentions or states and act in a way that optimizes performance. In realistic P300-speller simulations, AI outperforms traditional algorithms with an increase in bit rate between 18\% and 59\%, while offering a possibility of unifying various adaptive implementations within one generic framework.}
{Active Inference, BCI, P300 Speller, EEG}

\section{Introduction}

Adaptive BCIs have shown to improve performance \cite{mladenovic:hal-01542504}, however thorough adaptation is far from being reached, and a general and flexible framework to implement adaptive features is still lacking. We appeal to a generic Bayesian approach, called Active Inference (AI), which tightly couples perception and action \cite{friston2006free}. Endowing the machine with AI, enables: (1) to infer user's intentions or states by accumulating observations (e.g. electrophysiological data) in a flexible manner, as well as (2) to act adaptively in a way that optimizes performance. We illustrate AI applied to BCI using realistic P300-speller simulations. We demonstrate it can implement new features such as optimizing the sequence of flashed letters and yield significant bit rate increases.

\section{Material, Methods and Results}

Active Inference rests on an explicit probabilistic model of user and task. Key variables include observed data, user hidden states, and machine's action, as follows.

The observed data, here electroencephalagraphy (EEG) responses to target, non-target (P300 or not) and feedback stimuli (Error Potentials -- ErrPs or not), allow the machine to infer user's hidden states, here the intention to spell a letter or pause as well as the recognition of a target/non-target or a correct/incorrect feedback. Depending on the hidden states inferred, the computer has possible actions, here to flash in order to accumulate confidence about the target letter, to stop flashing and to display the chosen letter, or to switch off the screen if it infers an idle state of the user, i.e. no P300 response has been observed for some time.

Each hidden state is mapped onto observations through the data likelihood matrix which can be learned from calibration data. Given the machine's actions, the transitions between hidden states are modeled by a probability (Markov) martix. We also predefine the preference over all possible outcomes. Typically, the preferred outcome is to be in the state of observing a correctly spelled letter. Finally, a parameter $\gamma$ sets the exploration-exploitation tradeoff for action selection.

We compared AI to two classical approaches:

\begin{enumerate}
\item P300-spelling with a fixed flash number (12) of repetitions and pseudo-random flashing;
\item P300-spelling with pseudo-random flashing but optimal stopping \cite{mattout2015improving}.
\end{enumerate}

To do so, we used data from 18 subjects from a previous P300-speller experiment \cite{perrin2011detecting}. For each algorithm and subject, we simulated the spelling of 12000 letters.

Furthermore, to demonstrate AI's flexibility, we implemented a "LookAway" case, in which the machine would infer the user to be in idle state and would switch the screen off. We also simulated an ErrP classification enabling the automated detection of a wrongly spelled letter. In case of such detection, AI picks the next most probable letter to spell or choose to continue flashing to strengthen its confidence.

\begin{figure}[h!]
  \vspace{-1em} 
  \begin{center}
    \includegraphics[width=0.75\textwidth]{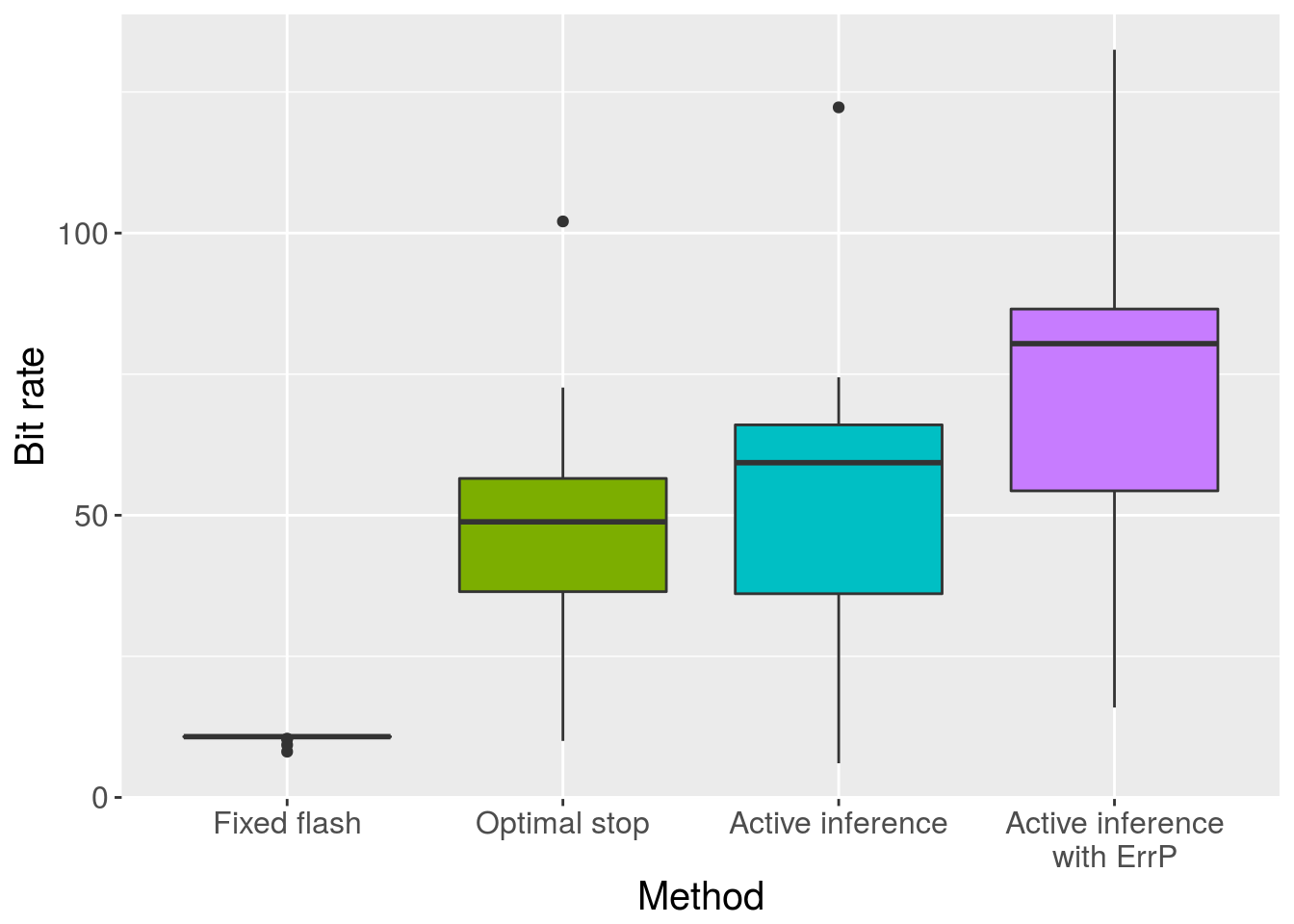}
    \vspace{-1em} 
    \caption{\textit{Comparison in bit rate (bit/min) between various flashing methods in a P300 BCI application. Data collected from the simulated spelling of 12000 letters with 18 subjects who were recorded in a previous experiment \cite{perrin2011detecting}. All methods significantly differ from one another (ANOVA, Tukey post-hoc, $p < 0.01$).}
    \label{fig:plot}}
  \end{center}
  \vspace{-1.5em} 
\end{figure}

AI showed significantly higher bit rate (54.12bit/min) than the second best strategy (optimal stopping, 45.70bit/min), see Figure \ref{fig:plot}. Its performance increased even further when a perfect ErrP classifier is used (73bit/min). Finally, when idle user states are simulated, it accurately switches off the speller $\approx$89\% of the time, after $\approx$24 flashes.

\section{Discussion}

Our results demonstrate a great potential for implementing adaptive BCI beyond existing approaches, showing an increase of 18\% and 59\% (using ErrP classifier) in bit rate.

\section{Significance}

AI outperforms other algorithms while offering a possibility of unifying various adaptive implementations within one generic framework. Thanks to such genericity, with only a few tuning of its parameters, AI can incorporate many features, such as automated correction or accounting for an idle user state. It can adjust to signal variability by inferring about the user, but it can also take into account the influence of its actions onto the user. This approach lays ground for future co-adaptive systems.


{\bibliography{biblio}}

\begin{thebibliography}{}

\bibitem[Friston et~al., 2006]{friston2006free}
Friston, K., Kilner, J., and Harrison, L. (2006).
\newblock A free energy principle for the brain.
\newblock {\em Journal of Physiology-Paris}, 100(1-3):70--87.

\bibitem[Mattout et~al., 2015]{mattout2015improving}
Mattout, J., Perrin, M., Bertrand, O., and Maby, E. (2015).
\newblock Improving bci performance through co-adaptation: Applications to the
  p300-speller.
\newblock {\em Annals of physical and rehabilitation medicine}, 58(1):23--28.

\bibitem[Mladenovi{\'c} et~al., 2017]{mladenovic:hal-01542504}
Mladenovi{\'c}, J., Mattout, J., and Lotte, F. (2017).
\newblock {A generic framework for adaptive EEG-based BCI training and
  operation}.
\newblock In Nam, C.~S., Nijholt, A., and Lotte, F., editors, {\em
  {Brain-Computer Interfaces Handbook: Technological and Theoretical
  Advances}}, volume~1 of {\em Brain-Computer Interfaces Handbook:
  Technological and Theoretical Advances}. {CRC Press: Taylor \& Francis
  Group}.

\bibitem[Perrin et~al., 2011]{perrin2011detecting}
Perrin, M., Maby, E., Bouet, R., Bertrand, O., and Mattout, J. (2011).
\newblock Detecting and interpreting responses to feedback in bci.
\newblock pages 116--119.

\end{thebibliography}

\end{document}